\documentclass[aps,pre,showpacs,notitlepage,amsmath,amssymb]{revtex4-1}
\usepackage{graphicx}

\begin{document}

\title{Anomalous diffusion in nonhomogeneous media: Power spectral density
of signals generated by time-subordinated nonlinear Langevin equations}

\author{R.~Kazakevi\v{c}ius}
\email{rytis.kazakevicius@tfai.vu.lt}

\affiliation{Institute of Theoretical Physics and Astronomy, Vilnius University,
A. Go\v{s}tauto 12, LT-01108 Vilnius, Lithuania}

\author{J.~Ruseckas}

\affiliation{Institute of Theoretical Physics and Astronomy, Vilnius University,
A. Go\v{s}tauto 12, LT-01108 Vilnius, Lithuania}

\begin{abstract}
Subdiffusive behavior of one-dimensional stochastic systems can be
described by time-subordinated Langevin equations. The corresponding
probability density satisfies the time-fractional Fokker-Planck equations.
In the homogeneous systems the power spectral density of the signals
generated by such Langevin equations has power-law dependency on the
frequency with the exponent smaller than 1. In this paper we consider
nonhomogeneous systems and show that in such systems the power spectral
density can have power-law behavior with the exponent equal to or
larger than 1 in a wide range of intermediate frequencies.
\end{abstract}


\maketitle

\section{Introduction}

A number of experimental observations show that more complex diffusion
processes in which the mean-square displacement is not proportional
to the time $t$ take place in various systems. A broad family of
processes described by certain deviations from the classical Brownian
linear time dependence of the centered second moment is called anomalous
diffusion. Anomalous diffusion in one dimension is characterized by
the occurrence of a mean square displacement of the form
\begin{equation}
\langle(\Delta x)^{2}\rangle=\frac{2K_{\alpha}}{\Gamma(1+\alpha)}t^{\alpha}\,,\label{eq:anom-diff}
\end{equation}
which deviates from the linear Brownian dependence on time \cite{Bouchaud1990}.
Eq.~(\ref{eq:anom-diff}) introduce the anomalous diffusion coefficient
$K_{\alpha}$. Such a deviation from classical diffusive behavior
can be observed in many systems \cite{Metzler2000,Scher2002,Golding2006}
and leads to many interesting physical properties \cite{Havlin1987}.
Applications of anomalous diffusion have been found in physics, chemistry
and biology \cite{Havlin1987,Isichenko1992,Bouchaud1990}. In general,
anomalous diffusion occurs in complex structures exhibiting the presence
of long-range correlations or memory effects \cite{Bouchaud1990}.
In the physics of complex systems, anomalous transport properties
and their description have attracted considerable interest starting
with the pioneering papers of Montroll and his collaborators \cite{Montroll1965}. 

An important subclass of anomalous diffusion processes constitute
subdiffusion processes, characterized by the sublinear dependence
with the power-law exponent in the range $0<\alpha<1$. In this situation
no finite mean jump time $\Delta t$ exists \cite{Metzler2000}. Subdiffusion
processes have been reported in condensed matter systems \cite{Metzler2000},
ecology \cite{Scher2002}, and biology \cite{Golding2006}. Continuous
time random walks (CTRWs) with on-site waiting-time distributions
falling slowly as $t^{-\alpha-1}$ and lacking the first moment predicts
a subdiffusive behavior and is a powerful tool to describe systems
which display subdiffusion \cite{Scher1975,Metzler2000}. Starting
from the generalized master equation or from the CTRW the fractional
Fokker-Planck equation can be rigorously derived \cite{Metzler1999a,Barkai2000}.
Fractional Fokker-Planck equation provides a useful approach for the
description of transport dynamics in complex systems which are governed
by anomalous diffusion \cite{Metzler2000} and nonexponential relaxation
patterns \cite{Jonscher2003}. It has been used to model dynamics
of protein systems and for reactions occurring in disordered media
\cite{Sabelko1999,Metzler1999b,Metzler2000,Sung2002,Seki2003,Metzler2004,Yuste2004,Chow2004}.
Description equivalent to a fractional Fokker-Planck equation consist
of a Markovian dynamics governed by an ordinary Langevin equation
but proceeding in an auxiliary, operational time instead of the physical
time \cite{Stanislavsky2003}. This Markovian process is subordinated
to the process defining the physical time; the subordinator introduces
memory effects \cite{Fogedby1994}. Other approaches for the theoretical
description of the subdiffusion use the generalized Langevin equation
\cite{Haenggi1990,Kou2004,Bao2006}, fractional Brownian motion \cite{Min2005},
or the Langevin equation with multiplicative noise \cite{Srokowski2009}.

The traditional CTRW provides a homogeneous description of the medium.
More complex situation is the diffusion in nonhomogeneous media, for
example diffusion on fractals and multifractals \cite{Schertzer2001}.
Nonhomogeneous systems exhibit not only subdiffusion related to traps,
but also enhanced diffusion can occur: for example, transport of interacting
particles in a weakly disordered media is superdiffusive due to the
disorder and subdiffusive without the disorder \cite{Ben-Naim2009}.
Anomalous diffusion in heterogeneous fractal medium has been considered
in Ref.~\cite{El-Wakil2001} where it was proposed that in one dimension
the mean square displacement has the form $\langle(\Delta x)^{2}\rangle\sim x^{-\theta}t^{\alpha}$
instead of Eq.~(\ref{eq:anom-diff}). Heterogeneous fractional Fokker\textendash{}Planck
equation on heterogeneous fractal structure media has been investigated
in Refs.~\cite{Ren2003,Ren2003a,Ren2003b,Qiu2004}. In nonhomogeneous
media the properties of a trap can reflect the medium structure, therefore
in the description of transport in such a medium the waiting time
should explicitly depend on the position. This dependence can be introduced
by using the position-dependent subdiffusion exponents \cite{Chechkin2005,Stickler2011,Stickler2011a}.
Another way is to consider position-dependent time subordinator \cite{Srokowski2014}.

In the homogeneous systems the power spectral density (PSD) of the
signals generated by time-subordinated Langevin equations has power-law
dependency $S(f)\sim f^{\alpha-1}$ on the frequency as $f\rightarrow0$.
\cite{Yim2006}. Since $0<\alpha<1$, the power-law exponent $1-\alpha$
is smaller than $1$. The purpose of this paper is to consider the
PSD in nonhomogeneous systems exhibiting anomalous diffusion. We demonstrate,
that in such systems the PSD can have power-law behavior with the
exponent equal to or larger than $1$ in a wide range of intermediate
frequencies.

The paper is organized as follows: In Sec.~\ref{sec:TFP} we introduce
the time-fractional Fokker-Planck equation describing subdiffusion
in nonhomogeneous media. The expression for the power spectral density
of the fluctuations of the diffusing particle in such a medium is
obtained in Sec.~\ref{sec:spectrum}. In Sec.~\ref{sec:power-law}
we consider a particular case of the time-fractional Fokker-Planck
equation involving the coefficients with power-law dependence on the
position. Numerical methods of solution are discussed in Sec.~\ref{sec:numer}.
Section~\ref{sec:concl} summarizes our findings.

\section{Time-fractional Fokker-Planck equation for nonhomogeneous media}

\label{sec:TFP}In this Section we derive the time-fractional Fokker-Planck
equation describing diffusion of a particle in nonhomogeneous media.
Usually the description of the anomalous diffusion is given by the
CTRW theory assuming heavy-tailed waiting-time distributions between
successive jumps of the diffusing particle. Here we use the method
of the derivation that is similar to that outlined in Refs.~\cite{Stanislavsky2003,Magdziarz2007}.
We start with the Markovian process described by the It\^o stochastic
differential equation (SDE)
\begin{equation}
dx(\tau)=a(x(\tau))d\tau+b(x(\tau))dW(\tau)\,.\label{eq:sde-1}
\end{equation}
Here $W(\tau)$ is the standard Brownian motion (Wiener process).
The drift coefficient $a(x)$ and the diffusion coefficient $b(x)$
explicitly depend on the particle position $x$. This dependence on
the position reflects the nonhomogeneity of a medium. Following Ref.~\cite{Stanislavsky2003}
we interpret the time $\tau$ in Eq.~(\ref{eq:sde-1}) as an internal,
operational time. Equation~(\ref{eq:sde-1}) we consider together
with an additional equation that relates the operational time $\tau$
to the physical time $t$. The difference between physical time $t$
and the operational time $\tau$ occurs due to trapping of the diffusing
particle. For the trapping processes that have distribution of the
trapping times with power law tails, the physical time $t=T(\tau)$
is given by the the strictly increasing $\alpha$-stable L\'evy motion
defined by the Laplace transform
\begin{equation}
\langle e^{-kT(\tau)}\rangle=e^{-\tau k^{\alpha}}\,.\label{eq:lapl-a-stable}
\end{equation}
Here the parameter $\alpha$ takes the values from the interval $0<\alpha<1$.
Thus the physical time $t$ obeys the SDE
\begin{equation}
dt(\tau)=dL^{\alpha}(\tau)\,,\label{eq:sde-t}
\end{equation}
where $dL^{\alpha}(\tau)$ stands for the increments of the strictly
increasing $\alpha$-stable L\'evy motion $L^{\alpha}(\tau)$. For
such physical time $t$ the operational time $\tau$ is related to
the physical time $t$ via the inverse $\alpha$-stable subordinator
\cite{Piryatinska2005,Magdziarz2006}
\begin{equation}
S(t)=\inf\{\tau:T(\tau)>t\}\,.\label{eq:inv-subord}
\end{equation}
The processes $x(\tau)$ and $S(t)$ are assumed to be independent.
Equations (\ref{eq:sde-1}) and (\ref{eq:sde-t}) define the subordinated
process $y(t)$ obtained by a random change of time
\begin{equation}
y(t)=x(S(t))\,.
\end{equation}
The process $y(t)$ describes the diffusion of a particle in a medium
with traps.

We will derive the equation for the probability density function (PDF)
of $y$. For the derivation we use the method of Laplace transform.
The PDF $P_{x}(x,\tau)$ of the stochastic variable $x$ as a function
of the operational time $\tau$ obeys the Fokker-Planck equation corresponding
to the It\^o SDE (\ref{eq:sde-1})
\begin{equation}
\frac{\partial}{\partial\tau}P_{x}(x,\tau)=L_{\mathrm{FP}}(x)P_{x}(x,\tau)\,,
\label{eq:FP}
\end{equation}
where $L_{\mathrm{FP}}(x)$ is the time-independent Fokker-Planck
operator \cite{Gardiner2004}
\begin{equation}
L_{\mathrm{FP}}(x)=-\frac{\partial}{\partial x}a(x)+\frac{1}{2}\frac{\partial^{2}}{\partial x^{2}}b^{2}(x)\,.\label{eq:LFP}
\end{equation}
The Laplace transform of Eq.~(\ref{eq:FP}) is
\begin{equation}
k\tilde{P}_{x}(x,k)-P_{x}(x,0)=L_{\mathrm{FP}}(x)\tilde{P}_{x}(x,k)\,.
\end{equation}
Since the processes $x(\tau)$ and $S(t)$ are independent, the PDF
of the random process $x(S(t))$ is given by
\begin{equation}
P(x,t)=\int P_{x}(x,\tau)P_{S}(\tau,t)\, d\tau\,.\label{eq:indep}
\end{equation}
Here $P_{S}(\tau,t)$ is the PDF of the inverse $\alpha$-stable subordinator
$S(t)$. From Eq.~(\ref{eq:indep}) it follows that the Laplace transform
$\tilde{P}(x,k)$ of the PDF $P(x,t)$ is related to the Laplace transform
$\tilde{P}_{S}(\tau,k)$ of the inverse subordinator $S(t)$:
\begin{equation}
\tilde{P}(x,k)=\int P_{x}(x,\tau)\tilde{P}_{S}(\tau,k)\, d\tau\,.\label{eq:lapl-y-lapl-s}
\end{equation}
The Laplace transform $\tilde{P}_{S}(\tau,k)$ of the inverse subordinator
$S(t)$ we obtain as follows: from the definition of the inverse subordinator
(\ref{eq:inv-subord}) we have $\mathrm{Pr}(S(t)<\tau)=\mathrm{Pr}(T(\tau)\geqslant t)$,
therefore
\begin{equation}
P_{S}(\tau,t)=-\frac{\partial}{\partial\tau}\int_{0}^{t}P_{T}(t',\tau)\, dt'\,.\label{eq:s-deriv-t}
\end{equation}
Here $P_{T}(t,\tau)$ is the PDF of the strictly increasing $\alpha$-stable
L\'evy motion $T(\tau)$. The PDF $P_{T}(t,\tau)$ fulfills the scaling
relation
\begin{equation}
P_{T}(t,\tau)=\frac{1}{\tau^{\frac{1}{\alpha}}}P_{T}\left(\frac{t}{\tau^{\frac{1}{\alpha}}},1\right)\,,\label{eq:scaling-u}
\end{equation}
since the strictly increasing $\alpha$-stable L\'evy motion is $1/\alpha$
self-similar \cite{Weron2005}. Combining Eqs.~(\ref{eq:s-deriv-t})
and (\ref{eq:scaling-u}) we obtain
\begin{equation}
P_{S}(\tau,t)=\frac{t}{\alpha\tau}P_{T}(t,\tau)\,.
\end{equation}
Consequently, the Laplace transform of $P_{S}(\tau,t)$ is equal to
\begin{equation}
\tilde{P}_{S}(\tau,k)=k^{\alpha-1}e^{-\tau k^{\alpha}}\,.\label{eq:lapl-s}
\end{equation}
Here we used Eq.~(\ref{eq:lapl-a-stable}) for the Laplace transform
of $P_{T}(t,\tau)$.

Using Eqs.~(\ref{eq:lapl-y-lapl-s}) and (\ref{eq:lapl-s}) we get
\begin{equation}
\tilde{P}(x,k)=k^{\alpha-1}\tilde{P}_{x}(x,k^{\alpha})\,.\label{eq:laplace-x}
\end{equation}
Acting with the operator $L_{\mathrm{FP}}(x)$ on Eq.~(\ref{eq:laplace-x})
we have
\begin{equation}
\tilde{P}(x,k)=k^{-1}P_{x}(x,0)+k^{-\alpha}L_{\mathrm{FP}}(x)\tilde{P}(x,k)\,.
\end{equation}
The inverse Laplace transform of this equation yields
\begin{equation}
P(x,t)=P_{x}(x,0)+\frac{1}{\Gamma(\alpha)}\int_{0}^{t}dt'\,(t-t')^{\alpha-1}L_{\mathrm{FP}}(x)P(x,t')\,.\label{eq:tmp1}
\end{equation}
Introducing the fractional Riemann-Liouville operator
\cite{Samko1993}
\begin{equation}
_{0}D_{t}^{-\alpha}f(t)\equiv\frac{1}{\Gamma(\alpha)}\int_{0}^{t}\frac{f(t')}{(t-t')^{1-\alpha}}dt'\,,\qquad0<\alpha<1
\end{equation}
we can write Eq.~(\ref{eq:tmp1}) as
\begin{equation}
P(x,t)=P_{x}(x,0)+{}_{0}D_{t}^{-\alpha}L_{\mathrm{FP}}(x)P(x,t)
\end{equation}
By differentiating this equation with respect to time we get the time-fractional
Fokker-Planck equation
\begin{equation}
\frac{\partial}{\partial t}P(x,t)={}_{0}D_{t}^{1-\alpha}\left(-\frac{\partial}{\partial x}[a(x)P]+\frac{1}{2}\frac{\partial^{2}}{\partial x^{2}}[b^{2}(x)P]\right)\,,\label{eq:time-fract-FP}
\end{equation}
where
\begin{equation}
_{0}D_{t}^{1-\alpha}f(t)\equiv\frac{1}{\Gamma(\alpha)}\frac{\partial}{\partial t}\int_{0}^{t}\frac{f(t')}{(t-t')^{1-\alpha}}dt'\,,\qquad0<\alpha<1
\end{equation}
The operator $_{0}D_{t}^{1-\alpha}$ is expressed via the convolution
with a slowly decaying kernel, which is typical for memory effects
in complex systems \cite{Stanislavsky2000}. Equation (\ref{eq:time-fract-FP})
is the equation describing the subdiffusion of particles in an inhomogeneous
medium. This equation generalizes the previously obtained time-fractional
Fokker-Planck equation with the position-independent diffusion coefficient.

\subsection{Position-dependent trapping time}

The properties of a trap in a nonhomogeneous medium can reflect the
structure of the medium. In the description of the transport in such
a medium the waiting time should explicitly depend on the position
\cite{Srokowski2014}. Instead of Eq.~(\ref{eq:sde-t}) we assume
that the physical time $t$ is related to the operational time $\tau$
via the SDE
\begin{equation}
dt(\tau)=g(x(\tau))dL^{\alpha}(\tau)\,.\label{eq:sde-t-2}
\end{equation}
Here the positive function $g(x)$ is the intensity of random time and models the
position of structures responsible for either trapping or accelerating the
particle. Large values of $g(x)$ corresponds to trapping of the particle,
whereas small $g(x)$ leads to the acceleration of diffusion. Similar equation
has been used in Ref.~\cite{Srokowski2014}. We interpret Eq.~(\ref{eq:sde-t-2})
according to the It\^o stochastic calculus: the values of $x$ and $t$ at
operational time $\tau$ are determined by events prior to the application of the
stochastic force $dL^{\alpha}$, which acts only from time $\tau$ to
$\tau+d\tau$. This assumption leads to the decoupling of the changes of $x$ and
the changes of $t$ occuring during an infinitesimal increment of the operational
time $d\tau$. Note, that the increments of the strictly increasing
$\alpha$-stable L\'evy motion $L^{\alpha}(\tau)$ are characterized by long tails
and thus only moments of order smaller than $\alpha$ are finite.

For fixed particle postion $x$ the coefficient $g(x)$ in Eq.~(\ref{eq:sde-t-2})
is constant and Eq.~(\ref{eq:sde-t-2}) corresponds to the fractional
Fokker-Planck equation
\begin{equation}
\frac{\partial}{\partial\tau}P(t;\tau|x)=
-{}_{0}D_{t}^{\alpha}g(x)^{\alpha}P(t;\tau|x)\,.
\label{eq:t-tau-tmp}
\end{equation}
This equation can be obtained by noting that from the definiton of the stricly
increasing $\alpha$-stable L\'evy motion (\ref{eq:lapl-a-stable}) the Laplace
transform of the PDF $P(t;\tau|x)$ is
$\tilde{P}(k;\tau|x)=\exp\{-\tau[g(x)k]^\alpha\}$. Diferentiating this
expression with respect to $\tau$ and taking the inverse Laplace transform one
gets Eq.~(\ref{eq:t-tau-tmp}). Alternatively, one can obtain 
the fractional Fokker-Planck equation using the methods of 
Refs.~\cite{Ditlevsen1999,Schertzer2001,Denisov2008}. The fractional derivative in the
Fokker-Planck equation appears as a consequence of the increments of L\'evy
$\alpha$-stable motion in Eq.~(\ref{eq:sde-t-2}).

Equations (\ref{eq:sde-1}) and (\ref{eq:sde-t-2}) together define the
subordinated process. However, now the processes $x(\tau)$ and $t(\tau)$ are not
independent and the derivation of the Fokker-Planck equation presented in
previous subsection is not applicable. Nevertheless, we can show that also with
position dependent trapping time the resulting equation has the form of
Eq.~(\ref{eq:time-fract-FP}). To do this let us consider the joint PDF
$P_{x,t}(x,t;\tau)$ of the stochastic variables $x$ and $t$.

SDEs (\ref{eq:sde-1}) and (\ref{eq:sde-t-2}) correspond to the following
two-dimensional fractional Fokker-Planck equation:
\begin{equation}
\frac{\partial}{\partial\tau}P_{x,t}(x,t;\tau)=L_{\mathrm{FP}}(x)P_{x,t}
-{}_{0}D_{t}^{\alpha}g(x)^{\alpha}P_{x,t}\,.
\label{eq:FP-x-t}
\end{equation}
This equation is a combination of Eqs.~(\ref{eq:FP}) and (\ref{eq:t-tau-tmp}).
Two-dimensional fractional Fokker-Planck equation (\ref{eq:FP-x-t}) for the PDF of two
stochastic variables $x$ and $t$ can be rigorously derived from the SDEs
(\ref{eq:sde-1}) and (\ref{eq:sde-t-2}) driven by L\'evy stable noises as in
Refs.~\cite{Ditlevsen1999,Schertzer2001,Denisov2008} (the Gaussian noise in
Eq.~(\ref{eq:sde-1}) is a particular case of a L\'evy stable noise with index of
stability $\alpha=2$).

The zero of the physical time $t$ coincides with the zero of the
operational time $\tau$, therefore, the initial condition for Eq.~(\ref{eq:FP-x-t})
is $P_{x,t}(x,t;0)=P_{x}(x,0)\delta(t)$. In addition, since $t$
is strictly increasing, we have a boundary condition $P_{x,t}(x,0;\tau)=0$
when $\tau>0$. The fractional Riemann-Liouville operator
$_{0}D_{t}^{\alpha}$ in Eq.~(\ref{eq:FP-x-t}) we can write as
$_{0}D_{t}^{\alpha}=\frac{\partial}{\partial t}{}_{0}D_{t}^{\alpha-1}$.

Now let us consider $x$ and $\tau$ as stochastic variables instead
of $x$ and $t$. Since the stochastic variable $t$ is related to
the operational time $\tau$ via Eq.~(\ref{eq:sde-t-2}), the joint
PDF $P_{x,\tau}(x,\tau;t)$ of the stochastic variables $x$ and $\tau$
is related to the PDF $P_{x,t}(x,t;\tau)$ according to the equation
\begin{equation}
P_{x,\tau}(x,\tau;t)={}_{0}D_{t}^{\alpha-1}g(x)^{\alpha}P_{x,t}(x,t;\tau)\,.\label{eq:p-x-tau}
\end{equation}
This equation can be obtained by noting that the last term in Eq.~(\ref{eq:FP-x-t})
contains derivative $\frac{\partial}{\partial t}$ and thus should
be equal to $-\frac{\partial}{\partial t}P_{x,\tau}$. From Eq.~(\ref{eq:p-x-tau})
if follows that 
\begin{equation}
P_{x,t}={}_{0}D_{t}^{1-\alpha}\frac{1}{g(x)^{\alpha}}P_{x,\tau}\,.\label{eq:p-x-t}
\end{equation}
Using Eqs.~(\ref{eq:FP-x-t}) and (\ref{eq:p-x-t}) we obtain
\begin{equation}
\frac{\partial}{\partial t}P_{x,\tau}(x,\tau;t)={}_{0}D_{t}^{1-\alpha}L_{\mathrm{FP}}(x)\frac{1}{g(x)^{\alpha}}P_{x,\tau}-\frac{\partial}{\partial\tau}{}_{0}D_{t}^{1-\alpha}\frac{1}{g(x)^{\alpha}}P_{x,\tau}\label{eq:FP-x-tau}
\end{equation}
The PDF $P_{x,\tau}$ has the initial condition $P_{x,\tau}(x,\tau;0)=P_{x}(x,0)\delta(\tau)$
and the boundary condition $P_{x,\tau}(x,0;t)=0$. The PDF of the
subordinated random process $x(t)$ is $P(x,t)=\int P_{x,\tau}(x,\tau;t)\, d\tau\,.$
Integrating both sides of Eq.~(\ref{eq:FP-x-tau}) we get
\begin{equation}
\frac{\partial}{\partial t}P(x,t)={}_{0}D_{t}^{1-\alpha}L_{\mathrm{FP}}^{\prime}(x)P\,,
\label{eq:FP-frac-2}
\end{equation}
where the new Fokker-Planck operator is 
\begin{equation}
L_{\mathrm{FP}}^{\prime}(x)=-\frac{\partial}{\partial x}a^{\prime}(x)
+\frac{1}{2}\frac{\partial^2}{\partial x^2}b^{\prime}(x)^{2}\,.
\end{equation}
Here the new drift and the diffusion coefficient are
\begin{equation}
a^{\prime}(x)=\frac{a(x)}{g(x)^{\alpha}}\,,\qquad b^{\prime}(x)=\frac{b(x)}{g(x)^{\frac{\alpha}{2}}}\,.\label{eq:new-coeff}
\end{equation}
Thus position-dependent trapping leads to position-dependent coefficients
in the time-fractional Fokker-Planck equation, even if the initial
SDE (\ref{eq:sde-1}) has constant coefficients. Eq.~(\ref{eq:FP-frac-2}) is the same as
Eq.~(\ref{eq:time-fract-FP}) when $g(x)$ is constant and does not depend on position.

\section{Power spectral density and time-fractional Fokker-Planck equation}

\label{sec:spectrum}In this Section we derive a general expression
for the PSD of the fluctuations of the diffusing particle in nonhomogeneous
medium. The evolution of the PDF of particle position $x$ is described
by the time-fractional Fokker-Planck equation (\ref{eq:time-fract-FP}).
For calculation of the spectrum we use the eigenfunction expansion
of the Fokker-Planck operator $L_{\mathrm{FP}}$. Method of eigenfunctions
for solving of time-dependent fractional Fokker-Planck equation has
been used in Ref.~\cite{Metzler1999}. Spectrum of fluctuations when
the diffusion coefficient is constant has been obtained in Ref.~\cite{Yim2006}.
Similar derivation of the spectrum for nonlinear SDE has been performed
in \cite{Ruseckas2010}.

The eigenfunctions of the Fokker-Planck operator $L_{\mathrm{FP}}(x)$
are the solutions of the equation
\begin{equation}
L_{\mathrm{FP}}(x)P_{\lambda}(x)=-\lambda P_{\lambda}(x)\,.\label{eq:eigen-general}
\end{equation}
Here $P_{\lambda}(x)$ are the eigenfunctions and $\lambda\geqslant0$
are the corresponding eigenvalues. The eigenfunctions obey the orthonormality
relation \cite{Risken1989}
\begin{equation}
\int e^{\Phi(x)}P_{\lambda}(x)P_{\lambda'}(x)dx=\delta_{\lambda,\lambda'}\,,
\end{equation}
where 
\begin{equation}
\Phi(x)=-\ln P_{0}(x)
\end{equation}
is the potential associated with the operator $L_{\mathrm{FP}}(x)$.
Here $P_{0}(x)$ is the steady-state solution of Eq.~(\ref{eq:time-fract-FP}).

We can write the time-dependent solution of the fractional Fokker-Planck
equation (\ref{eq:time-fract-FP}) corresponding to a single eigenfunction
as
\begin{equation}
P(x,t)=P_{\lambda}(x)f_{\lambda}(t)\,.\label{eq:x-t-sep}
\end{equation}
Inserting into Eq.~(\ref{eq:time-fract-FP}) we get that the function
$f(t)$ obeys the equation
\begin{equation}
\frac{d}{dt}f_{\lambda}(t)=-\lambda{}_{0}D_{t}^{1-\alpha}f_{\lambda}(t)\label{eq:ft}
\end{equation}
with the initial condition $f(0)=1$. The Laplace transform of this
equation yields
\begin{equation}
k\tilde{f}_{\lambda}(k)=1-\lambda k^{1-\alpha}\tilde{f}_{\lambda}(k)\,.\label{eq:ft-lapl}
\end{equation}
The solution of Eq.~(\ref{eq:ft-lapl}) is
\begin{equation}
\tilde{f}_{\lambda}(k)=\frac{1}{k+\lambda k^{1-\alpha}}\,.\label{eq:ft-lapl-sol}
\end{equation}
The inverse Laplace transform is given in terms of the monotonically
decreasing Mittag-Leffler function \cite{Metzler1999}
\begin{equation}
f_{\lambda}(t)=E_{\alpha}(-\lambda t^{\alpha})\,.\label{eq:f-ml}
\end{equation}
The Mittag-Leffler function has a series expansion 
\begin{equation}
E_{\alpha}(z)\equiv E_{\alpha,1}(z)=\sum_{n=0}^{\infty}\frac{z^{n}}{\Gamma(\alpha n+1)}\,.
\end{equation}

The autocorrelation function can be calculated from the transition
probability $P(x,t|x_{0},0)$ (the conditional probability that at
time $t$ the stochastic variable has value $x$ with the condition
that at time $t=0$ it had the value $x_{0}$):
\begin{equation}
C(t)=\int dx\int dx_{0}\, x_{0}xP_{0}(x_{0})P(x,t|x_{0},0)-\left[\int dx\, xP_{0}(x)\right]^{2}\label{eq:autocorr}
\end{equation}
The transition probability is the solution of the Fokker-Planck equation
(\ref{eq:time-fract-FP}) with the initial condition $P(x,0|x_{0},0)=\delta(x-x_{0})$.
Expansion of the transition probability density in a series of the
eigenfunctions has the form
\begin{equation}
P(x,t|x_{0},0)=\sum_{\lambda}P_{\lambda}(x)e^{\Phi(x_{0})}P_{\lambda}(x_{0})E_{\alpha}(-\lambda t^{\alpha})\,,\label{eq:trans-exp}
\end{equation}
where we used Eqs.~(\ref{eq:x-t-sep}) and (\ref{eq:f-ml}). Inserting
Eq.~(\ref{eq:trans-exp}) into Eq.~(\ref{eq:autocorr}) we get the
expression for the autocorrelation function
\begin{equation}
C(t)=\sum_{\lambda>0}X_{\lambda}^{2}E_{\alpha}(-\lambda t^{\alpha})\,.\label{eq:autocorr-2}
\end{equation}
Here
\begin{equation}
X_{\lambda}=\int xP_{\lambda}(x)\, dx
\end{equation}
is the first moment of the stochastic variable $x$ evaluated with
the $\lambda$-th eigenfunction $P_{\lambda}(x)$. Such an expression
for the autocorrelation function has been obtained in Ref.~\cite{Yim2006}.

According to Wiener-Khintchine relations, the power spectral density
is related to the autocorrelation function: 
\begin{equation}
S(f)=4\int_{0}^{\infty}C(t)\cos(\omega t)\, dt\,,
\end{equation}
where $\omega=2\pi f$. Using Eq.~(\ref{eq:autocorr-2}) we obtain
\begin{equation}
S(f)=4\sum_{\lambda>0}X_{\lambda}^{2}\int_{0}^{\infty}E_{\alpha}(-\lambda t^{\alpha})\cos(\omega t)\, dt
\end{equation}
The integral can be calculated by noting that the Laplace transform
of $E_{\alpha}(-\lambda t^{\alpha})$ is given by Eq.~(\ref{eq:ft-lapl-sol}).
We obtain the desired expression for the PSD
\begin{equation}
S(f)=4\frac{\sin\left(\frac{\pi}{2}\alpha\right)}{\omega^{1-\alpha}}\sum_{\lambda}\frac{\lambda}{\lambda^{2}+\omega^{2\alpha}+2\lambda\omega^{\alpha}\cos\left(\frac{\pi}{2}\alpha\right)}X_{\lambda}^{2}\,.\label{eq:psd-gen}
\end{equation}
Eq.~(\ref{eq:psd-gen}) becomes the usual expression for the PSD
when $\alpha\rightarrow1$. Similar expression for the spectrum has
been obtained in Ref.~\cite{Yim2006}.

For small frequencies $\omega\ll\lambda_{1}^{1/\alpha}$ we can neglect
the frequency when it appears together with the eigenvalues $\lambda$.
Here $\lambda_{1}$ is the smallest eigenvalue larger than zero. Thus
for small frequencies Eq.~(\ref{eq:psd-gen}) approximately is
\begin{equation}
S(f)\approx4\frac{\sin\left(\frac{\pi}{2}\alpha\right)}{\omega^{1-\alpha}}\sum_{\lambda}\frac{X_{\lambda}^{2}}{\lambda}\,.\label{eq:psd-small}
\end{equation}
We obtain that for small frequencies the PSD has a power-law dependency
on the frequency $S(f)\sim f^{-(1-\alpha)}$. However, the power-law
exponent is always smaller than $1$, since $0<\alpha<1$. It is not
possible to get pure $1/f$ spectrum this way. In the next Section
we show that it is possible to get larger power-law exponents in the
PSD in a wide range of intermediate frequencies when the diffusion
coefficient is not constant and depends on $x$.

\section{Time-fractional Fokker-Planck equation with power-law coefficients}

\label{sec:power-law}In this Section we consider a particular case
of the time-fractional Fokker-Planck equation (\ref{eq:time-fract-FP}).
We assume that the diffusion coefficient has a power-law dependence
on the particle position $x$ and Eq.~(\ref{eq:time-fract-FP}) takes
the form
\begin{equation}
\frac{\partial}{\partial t}P(x,t)=\sigma^{2}{}_{0}D_{t}^{1-\alpha}
\left\{ \left(\frac{\nu}{2}-\eta\right)
\frac{\partial}{\partial x}\left[x^{2\eta-1}P(x,t)\right]
+\frac{1}{2}\frac{\partial^{2}}{\partial x^{2}}\left[x^{2\eta}P(x,t)\right]\right\} \,.
\label{eq:FP-frac}
\end{equation}
Here $\eta$ is the power-law exponent of the multiplicative noise in
Eq.~(\ref{eq:sde-1}) and $\nu$ defines the behavior of the steady-state PDF
$P_{0}(x)$. Eq.~(\ref{eq:FP-frac}) should be considered together with the
boundary conditions that restrict the stochastic variable $x$ to the positive
values.

The steady-state PDF $P_{0}(x)$ obtained from Eq.~(\ref{eq:FP-frac})
has a power-law form
\begin{equation}
P_{0}(x)\sim x^{-\nu}\,.\label{eq:p-steady}
\end{equation}
For $\nu\ge 1$ the PDF $P_{0}(x)$ diverges as $x\rightarrow0$, thus the diffusion
should be restricted at least from the side of small values. This can be done by
introducing an additional potential that becomes large only when $x$ acquires
values outside of the interval $[x_{\mathrm{min}},x_{\mathrm{max}}]$ into the
drift term of Eq.~(\ref{eq:FP-frac}). The simplest choice is the reflective
boundaries at $x=x_{\mathrm{min}}$ and $x=x_{\mathrm{max}}$.

The power-law form of the diffusion coefficient is natural for systems
exhibiting self-similarity, for example disordered materials, and has been used
to describe diffusion on fractals \cite{OShaughnessy1985,Metzler1994}, turbulent
two-particle diffusion, transport of fast electrons in a hot plasma
\cite{Vedenov1967,Fujisaka1985}. Equation~(\ref{eq:FP-frac}) is a generalization
of the Fokker-Planck equation resulting form nonlinear SDEs proposed in
Refs.~\cite{Kaulakys2004,Kaulakys2006}. Such nonlinear SDEs generate signals
having $1/f$ spectrum in a wide range of frequencies and have been used to
describe signals in socio-economical systems \cite{Gontis2010,Mathiesen2013} and
Brownian motion in inhomogeneous media \cite{Kazakevicius2014}.

In Ref.~\cite{Ruseckas2010} an approximate expression for the first
moment $X_{\lambda}$ has been obtained for the Fokker-Planck operator
appearing in Eq.~(\ref{eq:FP-frac}) assuming reflective boundaries
at $x_{\mathrm{min}}=1$ and $x_{\mathrm{max}}=\xi$, $\xi\gg1$.
According to the results of Ref.~\cite{Ruseckas2010} 
\begin{equation}
X_{\lambda}\sim\frac{c_{\lambda}}{|1-\eta|}\frac{1}{\rho^{\beta_{1}}}\,,\label{eq:x-lambda}
\end{equation}
where
\begin{equation}
c_{\lambda}=\sqrt{\frac{|1-\eta|}{z_{\mathrm{max}}}\frac{\nu-1}{1-\xi^{1-\nu}}\pi\rho}\,,\qquad\rho=\frac{\sqrt{2\lambda}}{|\eta-1|}\,,\qquad\beta_{1}=1+\frac{\nu-3}{2(\eta-1)}\,.
\end{equation}
The parameters $z_{\mathrm{min}}$ and $z_{\mathrm{max}}$ depend
on the boundaries $x_{\mathrm{min}}$ and $x_{\mathrm{max}}$. When
$\rho z_{\mathrm{max}}\gg1$, replacing summation by integration in
Eq.~(\ref{eq:psd-gen}) we obtain the expression for the PSD
\begin{equation}
S(f)\approx4\frac{\sin\left(\frac{\pi}{2}\alpha\right)}{\omega^{1-\alpha}}\int\frac{\lambda}{\lambda^{2}+\omega^{2\alpha}+2\lambda\omega^{\alpha}\cos\left(\frac{\pi}{2}\alpha\right)}X_{\lambda}^{2}D(\lambda)\, d\lambda
\end{equation}
The density of eigenvalues $D(\lambda)$ has been estimated as \cite{Ruseckas2010}
\begin{equation}
D(\lambda)\sim\frac{1}{\sqrt{\lambda}}\,.\label{eq:d-lambda}
\end{equation}
Using Eqs.~(\ref{eq:x-lambda}) and (\ref{eq:d-lambda}) we get
\begin{equation}
S(f)\sim4\frac{\sin\left(\frac{\pi}{2}\alpha\right)}{\omega^{1+\alpha(\beta_{1}-1)}}\int_{\frac{z_{\mathrm{max}}^{-2}}{\omega^{\alpha}}}^{\frac{z_{\mathrm{min}}^{-2}}{\omega^{\alpha}}}\frac{1}{u^{\beta_{1}-1}}\frac{1}{\left(u^{2}+1+2u\cos\left(\frac{\pi}{2}\alpha\right)\right)}du
\end{equation}
Here the upper range of integration is limited because $X_{\lambda}$
becomes small when $\rho z_{\mathrm{min}}\gg1$ \cite{Ruseckas2010}.
When $z_{\mathrm{max}}^{-2}\ll\omega^{\alpha}\ll z_{\mathrm{min}}^{-2}$
and $0<\beta_{1}<2$ then we can approximate the lower limit of integration
by $0$ and the upper limit by $\infty$. In this case the PSD depends
on the frequency as $S(f)\sim f^{-1-\alpha(\beta_{1}-1)}$. When $\beta_{1}>2$
then the largest contribution is from the lower limit of the integration.
Thus, when $z_{\mathrm{max}}^{-2}\ll\omega^{\alpha}\ll z_{\mathrm{min}}^{-2}$
then the leading term in the expansion of the approximate expression
for the PSD in the power series of $\omega$ is
\begin{equation}
S(f)\sim\begin{cases}
\frac{1}{\omega^{1+\alpha(\beta_{1}-1)}}\,, & 0<\beta_{1}<2\,,\\
\frac{1}{\omega^{1+\alpha}}\,, & \beta_{1}>2\,.
\end{cases}
\end{equation}
This expressions for PSD can also be written as
\begin{equation}
S(f)\sim\begin{cases}
\frac{1}{\omega^{\beta}}\,, & 1-\alpha<\beta<1+\alpha\,,\\
\frac{1}{\omega^{1+\alpha}}\,, & \beta>1+\alpha\,.
\end{cases}
\end{equation}
Here
\begin{equation}
\beta=1+\alpha(\beta_{1}-1)=1+\frac{\alpha(\nu-3)}{2(\eta-1)}\label{eq:beta}
\end{equation}
is the power-law exponent of the PSD. Equation (\ref{eq:beta}) generalizes
the expression for the power-law exponent obtained for nonlinear SDEs
\cite{Kaulakys2006}. When $\nu=3$ then from Eq.~(\ref{eq:beta})
follows that we obtain $1/f$ spectrum.

\subsection{Power spectral density from scaling properties}

Power-law exponent (\ref{eq:beta}) in the PSD can be obtained from
the scaling properties of Eq.~(\ref{eq:FP-frac}), similarly as it
has been done for the nonlinear SDEs \cite{Ruseckas2014}. Changing
the variable $x$ to the scaled variable $x_{s}=ax$ in Eq.~(\ref{eq:FP-frac})
yields
\begin{equation}
\frac{\partial}{\partial t}P(x_{s},t)=\frac{\sigma^{2}}{a^{2(\eta-1)}}{}_{0}D_{t}^{1-\alpha}\left\{ \left(\frac{\lambda}{2}-\eta\right)\frac{\partial}{\partial x_{s}}\left[x_{s}^{2\eta-1}P(x_{s},t)\right]+\frac{1}{2}\frac{\partial^{2}}{\partial x_{s}^{2}}\left[x_{s}^{2\eta}P(x_{s},t)\right]\right\} \,.
\end{equation}
The Riemann\textendash{}Liouville fractional derivative has the following
scaling property: $_{0}D_{t}^{1-\alpha}f(ct)=c^{1-\alpha}{}_{0}D_{ct}^{1-\alpha}f(ct)$.
Thus, changing the time $t$ to the scaled time $t_{s}=a^{\mu}t$
we get
\begin{equation}
a^{\mu}\frac{\partial}{\partial t_{s}}P(x,t_{s})=\sigma^{2}{}_{0}a^{\mu(1-\alpha)}D_{t_{s}}^{1-\alpha}\left\{ \left(\frac{\lambda}{2}-\eta\right)\frac{\partial}{\partial x}\left[x^{2\eta-1}P(x,t_{s})\right]+\frac{1}{2}\frac{\partial^{2}}{\partial x^{2}}\left[x^{2\eta}P(x,t_{s})\right]\right\} \,.
\end{equation}
The change of the variable $x$ to the scaled variable $ax$ or the
change of the time $t$ to the scaled time $a^{\mu}t$ produce the
same fractional Fokker-Planck equation if
\begin{equation}
\mu=\frac{2(\eta-1)}{\alpha}\,.\label{eq:mu}
\end{equation}
It follows, that the transition probability $P(x,t|x_{0},0)$ has
the following scaling property:
\begin{equation}
aP(ax,t|ax_{0},0)=P(x,a^{\mu}t|x_{0},0)\,.\label{eq:trans-prob-scaling}
\end{equation}
As has been shown in Ref.~\cite{Ruseckas2014}, the power-law steady
state PDF $P_{0}(x)\sim x^{-\nu}$ and the scaling property of the
transition probability (\ref{eq:trans-prob-scaling}) lead to the
power-law form PSD $S(f)\sim f^{-\beta}$ in a wide range of frequencies.
The power-law exponent $\beta$ is given by
\begin{equation}
\beta=1+(\nu-3)/\mu\,.
\end{equation}
Using Eq.~(\ref{eq:mu}) we obtain the same expression for $\beta$
as in Eq.~(\ref{eq:beta}).

The presence of restrictions at $x=x_{\mathrm{min}}$ and $x=x_{\mathrm{max}}$
makes the scaling (\ref{eq:trans-prob-scaling}) not exact. This limits
the power-law part of the PSD to a finite range of frequencies $f_{\mathrm{min}}\ll f\ll f_{\mathrm{max}}$.
Similarly as in Ref.~\cite{Ruseckas2014}, we estimate the limiting
frequencies as
\begin{eqnarray}
\sigma^{\frac{2}{\alpha}}x_{\mathrm{min}}^{\frac{2}{\alpha}(\eta-1)} & \ll & 2\pi f\ll\sigma^{\frac{2}{\alpha}}x_{\mathrm{max}}^{\frac{2}{\alpha}(\eta-1)}\,,\qquad\eta>1\,,\\
\sigma^{\frac{2}{\alpha}}x_{\mathrm{max}}^{-\frac{2}{\alpha}(1-\eta)} & \ll & 2\pi f\ll\sigma^{\frac{2}{\alpha}}x_{\mathrm{min}}^{-\frac{2}{\alpha}(1-\eta)}\,,\qquad\eta<1\,.\nonumber 
\end{eqnarray}
This equation shows that the frequency range grows with decrease of
$\alpha$. By increasing the ratio $x_{\mathrm{max}}/x_{\mathrm{min}}$
one can get an arbitrarily wide range of the frequencies where the
PSD has $1/f^{\beta}$ behavior.

\section{\label{sec:numer}Numerical approach}

\subsection{Numerical approximation of sample paths}

Since analytical solution of time-fractional Fokker-Planck equation
can be obtained only in separate cases, there is a need of numerical
solution. Numerical solution of time-fractional Fokker-Planck equation
is complicated \cite{Deng2007}. It is easier to numerically solve
Langevin equations (\ref{eq:sde-1}), (\ref{eq:sde-t}) instead. The
desired properties of the solution of the Fokker-Planck equation then
can be calculated by averaging over many sample paths obtained by
solving the Langevin equations. The numerical method of solution of
the Langevin equations with constant drift coefficient is outlined
in \cite{Magdziarz2007,Gajda2010}. We can use the same method also
when the drift coefficient is position-dependent.

Choosing the time step $\Delta\tau$ of the operational time $\tau$
the inverse subordinator $S(t)$ is approximated as \cite{Magdziarz2009}
\begin{equation}
S_{\Delta\tau}(t)=[\min\{n\in\mathbb{N}:T(n\Delta\tau)>t\}-1]\Delta\tau\,.
\end{equation}
Such approximation satisfies \cite{Magdziarz2009a}
\begin{equation}
\sup_{0\leqslant t\leqslant T}[S_{\Delta\tau}(t)-S(t)]\leqslant\Delta\tau\,.
\end{equation}
The values $T(n\Delta\tau)$ are generated by summing up the independent
and stationary increments of the L\'evy process:
\begin{equation}
T(n\Delta\tau)=T([n-1]\Delta\tau)+\Delta\tau^{1/\alpha}\xi_{n}\,.\label{eq:t-appr}
\end{equation}
Here $\xi_{n}$ are independent totally skewed positive $\alpha$-stable
random variables with the distribution specified by the Laplace transform
$\langle e^{-k\xi}\rangle=e^{-k^{\alpha}}$ . Such variables can be
generated using the formula \cite{Weron1996} 
\begin{equation}
\xi=\frac{\sin\left[\alpha\left(U+\frac{\pi}{2}\right)\right]}{\cos(U)^{\frac{1}{\alpha}}}\left(\frac{\cos\left[U-\alpha\left(U+\frac{\pi}{2}\right)\right]}{W}\right)^{\frac{1-\alpha}{\alpha}}\,.
\end{equation}
Here $U$ is uniformly distributed on $\left(-\frac{\pi}{2},\frac{\pi}{2}\right)$
and $W$ has an exponential distribution with mean $1$. Note, that
in Ref.~\cite{Magdziarz2007} incorrect formula for generating totally
skewed positive $\alpha$-stable random variables has been used. The
definition of the L\'evy $\alpha$-stable distribution using the
Laplace transform (\ref{eq:lapl-a-stable}) differs from the more
common definition using the Fourier transform. This has been corrected
in Ref.~\cite{Gajda2010}.

The SDE (\ref{eq:sde-1}) in the operational time $\tau$ can be numerically
solved using the Euler-Maruyama scheme with the time step $\Delta\tau$.
For each value of the stochastic variable $x_{k}$ we assign the physical
time $t_{k}$ generated by the process $T(\tau)$ using Eq.~(\ref{eq:t-appr}).
Thus the numerical method of solution of Langevin equations (\ref{eq:sde-1}),
(\ref{eq:sde-t}) is given by the following equations:
\begin{eqnarray}
x_{k+1} & = & x_{k}+a(x_{k})\Delta\tau+b(x_{k})\sqrt{\Delta\tau}\varepsilon_{k}\,,\label{eq:numer-1}\\
t_{k+1} & = & t_{k}+\Delta\tau^{\frac{1}{\alpha}}\xi_{k}\,.\label{eq:numer-2}
\end{eqnarray}
Here $\varepsilon_{k}$ are i.i.d. random variables having standard
normal distribution.

For numerical solution of nonlinear equations, such as those resulting
in Eq.~(\ref{eq:FP-frac}), the fixed time step $\Delta\tau$ can
be inefficient. For example, in Eq.~(\ref{eq:FP-frac}) with $\eta>1$
large values of stochastic variable $x$ lead to large coefficients
and thus require a very small time step. A more efficient way of solution
is to use a variable time step that adapts to the coefficients in
the equation. Similar method has been used in Refs.~\cite{Kaulakys2004,Kaulakys2006}
for solving nonlinear SDEs. Such a variable time step is equivalent
to changing of the operational time $\tau$ to the position-dependent
operational time $\tau^{\prime}$. If we choose the intensity of random
time in Eq.~(\ref{eq:sde-t-2}) as $g(x)=b(x)^{-\frac{2}{\alpha}}$
then, according to Eq.~(\ref{eq:new-coeff}) instead of initial Langevin
equations (\ref{eq:sde-1}), (\ref{eq:sde-t}) we get the new Langevin
equations
\begin{eqnarray}
dx(\tau^{\prime}) & = & \frac{a(x(\tau^{\prime}))}{b(x(\tau^{\prime}))^{2}}+dW(\tau^{\prime})\,,\label{eq:sde-3}\\
dt(\tau^{\prime}) & = & b(x(\tau^{\prime}))^{-\frac{2}{\alpha}}dL^{\alpha}(\tau^{\prime})\,.
\end{eqnarray}
Discretizing the operational time $\tau^{\prime}$ with the time step
$\Delta\tau^{\prime}$ and using the Euler-Maruyama approximation
for Eq.~(\ref{eq:sde-3}) instead of Eqs.~(\ref{eq:numer-1}), (\ref{eq:numer-2})
we have
\begin{eqnarray}
x_{k+1} & = & x_{k}+\frac{a(x_{k})}{b(x_{k})^{2}}\Delta\tau^{\prime}+\sqrt{\Delta\tau^{\prime}}\varepsilon_{k}\,,\label{eq:numer-3}\\
t_{k+1} & = & t_{k}+\left(\frac{\Delta\tau^{\prime}}{b(x_{k})^{2}}\right)^{\frac{1}{\alpha}}\xi_{k}\,.\label{eq:numer-4}
\end{eqnarray}
Comparison with Eqs.~(\ref{eq:numer-1}), (\ref{eq:numer-2}) shows
that Eqs.~(\ref{eq:numer-3}), (\ref{eq:numer-4}) can be obtained
by replacing the time step $\Delta\tau$ in Eqs.~(\ref{eq:numer-1}),
(\ref{eq:numer-2}) by 
\begin{equation}
\Delta\tau\rightarrow\frac{\Delta\tau^{\prime}}{b(x_{k})^{2}}\,.
\end{equation}

As an example, we solve the Langevin equations
\begin{eqnarray}
dx & = & \left(\eta-\frac{\nu}{2}\right)x^{2\eta-1}d\tau+x^{\eta}dW(\tau)\,,\label{eq:example-1}\\
dt & = & dL^{\alpha}(\tau)\label{eq:example-2}
\end{eqnarray}
resulting in the time-fractional Fokker-Planck equation (\ref{eq:FP-frac}).
For restriction of the diffusion region we use the reflective boundaries
at $x=x_{\mathrm{min}}$ and $x_{\mathrm{max}}$. More effective numerical
solution scheme is obtained changing the operational time $\tau$
to the time $\tau'$ defined by the equation
\begin{equation}
dt(\tau^{\prime})=x(\tau^{\prime})^{-\frac{2}{\alpha}(\eta-1)}dL^{\alpha}(\tau^{\prime})\,.\label{eq:example-3}
\end{equation}
This change is equivalent to the introduction of the variable time
step $\Delta\tau_{k}=\Delta\tau^{\prime}x_{k}^{-2(\eta-1)}$. Discretizing
the operational time $\tau^{\prime}$ with the step $\Delta\tau^{\prime}$
from Eqs.~(\ref{eq:example-1})--(\ref{eq:example-3}) we get the
following numerical approximation:
\begin{eqnarray}
x_{k+1} & = & x_{k}+\left(\eta-\frac{\nu}{2}\right)x_{k}\Delta\tau^{\prime}+x_{k}\sqrt{\Delta\tau^{\prime}}\varepsilon_{k}\,,\label{eq:numer-5}\\
t_{k+1} & = & t_{k}+\left(\frac{\Delta\tau^{\prime}}{x_{k}^{2(\eta-1)}}\right)^{\frac{1}{\alpha}}\xi_{k}\,.\label{eq:numer-6}
\end{eqnarray}
\begin{figure}
\includegraphics[width=0.45\textwidth]{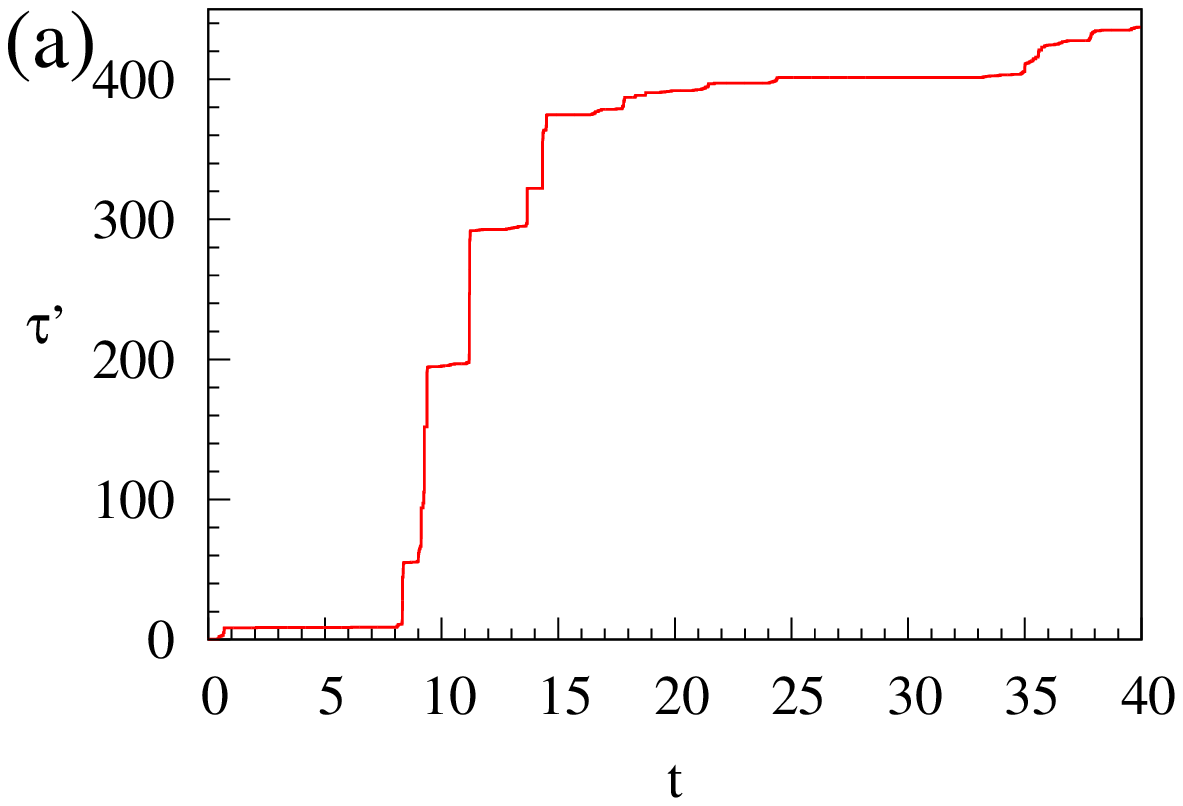}\includegraphics[width=0.45\textwidth]{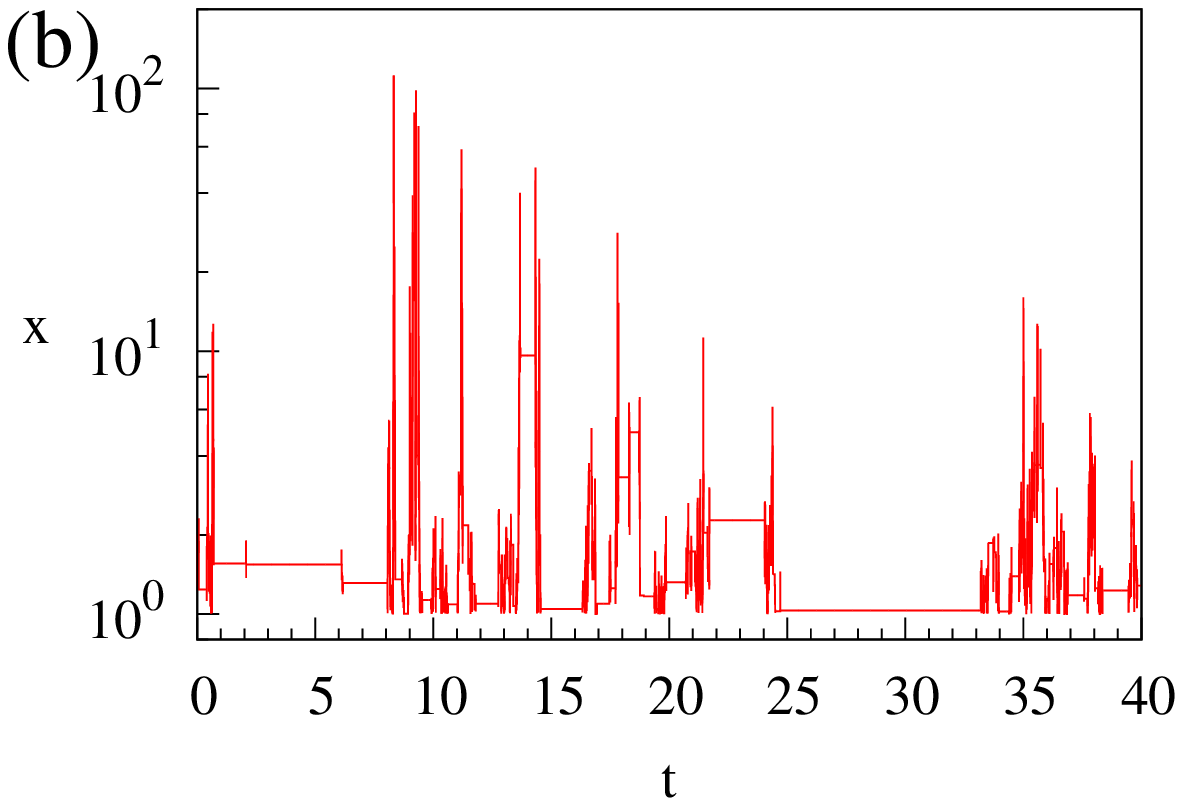}

\caption{Sample path obtained from Langevin equations (\ref{eq:example-1}),
(\ref{eq:example-2}) using numerical solution scheme given by Eqs.~(\ref{eq:numer-5}),
(\ref{eq:numer-6}). (a) Dependence of the operational time $\tau^{\prime}$,
defined by Eq.~(\ref{eq:example-3}), on the physical time $t$.
(b). Dependence of the stochastic variable $x$ on the physical time
$t$. The parameters are $\alpha=0.7$, $\eta=2$, $\nu=3$. Reflective
boundaries are placed at $x_{\mathrm{min}}=1$ and $x_{\mathrm{max}}=1000$.}
\label{fig:sample-path}
\end{figure}

Sample path obtained using Eqs.~(\ref{eq:numer-5}), (\ref{eq:numer-6})
with the parameters $\eta=2$ and $\nu=3$ is shown in Fig.~\ref{fig:sample-path}.
The change of the operational time $\tau'$ with the physical time
$t$ is shown in Fig.~\ref{fig:sample-path}(a) and the dependence
of the stochastic variable $x$ on the physical time $t$ is shown
in Fig.~\ref{fig:sample-path}(b). Due to nonlinear coefficients
in Eq.~(\ref{eq:example-1}) the sample path in Fig.~\ref{fig:sample-path}(b)
exhibits peaks or bursts, corresponding to the large deviations of
the variable $x$. The intervals with $x$ being constant indicate
the heavy-tailed trapping times. Comparing Fig.~\ref{fig:sample-path}(a)
with Fig.~\ref{fig:sample-path}(b) we see that the operational time
$\tau'$ increases faster when $x$ acquires larger values, in accordance
to Eq.~(\ref{eq:example-3}).

\subsection{Power spectral density}

Since the equations exhibit a slow (power-law instead of a usual exponential)
relaxation \cite{Metzler1999}, calculation of the PSD using sample
paths is very slow. More efficient way is to find the eigenvalues
and eigenfunctions of the Fokker-Planck operator (\ref{eq:LFP}) and
calculate the PSD using the rapidly converging series in Eq.~(\ref{eq:psd-gen}).
This is the approach for calculating the PSD used in Ref.~\cite{Yim2006}
for the case of constant diffusion coefficient.

As an example let us calculate the PSD of the diffusion described
by the time-fractional Fokker-Planck equation (\ref{eq:FP-frac})
with $\eta\neq1$ and the reflective boundaries at $x_{\mathrm{min}}=1$
and $x_{\mathrm{max}}=\xi$. The equation (\ref{eq:eigen-general})
for the eigenfunctions of the Fokker-Planck operator that enters Eq.~(\ref{eq:FP-frac})
is
\begin{equation}
-\left(\eta-\frac{\nu}{2}\right)\frac{\partial}{\partial x}x^{2\eta-1}P_{\lambda}(x)+\frac{1}{2}\frac{\partial^{2}}{\partial x^{2}}x^{2\eta}P_{\lambda}(x)=-\lambda P_{\lambda}(x)\,.\label{eq:eigen-1}
\end{equation}
The reflective boundaries lead to the conditions $S_{\lambda}(1)=0$
and $S_{\lambda}(\xi)=0$, where
\begin{equation}
S_{\lambda}(x)=\left(\eta-\frac{\nu}{2}\right)x^{2\eta-1}P_{\lambda}(x)-\frac{1}{2}\frac{\partial}{\partial x}x^{2\eta}P_{\lambda}(x)
\end{equation}
is the probability current related to the eigenfunction $P_{\lambda}(x)$.
The steady state solution of Eq.~(\ref{eq:FP-frac}) is
\begin{equation}
P_{0}(x)=\frac{\nu-1}{1-\xi^{1-\nu}}x^{-\nu}\,.
\end{equation}
It is more convenient to transform Eq.~(\ref{eq:eigen-1}) into the
Schr\"odinger equation \cite{Risken1989}. To do this we first make
the diffusion coefficient constant by changing the variable $x$ to
\begin{equation}
z=\frac{x^{1-\eta}}{|\eta-1|}\,.\label{eq:transformation}
\end{equation}
Eq.~(\ref{eq:eigen-1}) then becomes
\begin{equation}
\frac{\nu'}{2}\frac{\partial}{\partial z}\frac{1}{z}P_{\lambda}^{\prime}(z)+\frac{1}{2}\frac{\partial^{2}}{\partial z^{2}}P_{\lambda}^{\prime}(z)=-\lambda P_{\lambda}^{\prime}(z)\label{eq:eigen-2}
\end{equation}
with the reflective boundaries at $z_{\mathrm{min}}$ and $z_{\mathrm{max}}$,
where
\begin{equation}
z_{\mathrm{min}}=\begin{cases}
\frac{1}{\eta-1}\frac{1}{\xi^{\eta-1}}\,, & \eta>1\,,\\
\frac{1}{1-\eta}\,, & \eta<1\,,
\end{cases}\qquad z_{\mathrm{max}}=\begin{cases}
\frac{1}{\eta-1}\,, & \eta>1\,,\\
\frac{1}{1-\eta}\xi^{1-\eta}\,, & \eta<1\,.
\end{cases}
\end{equation}
Here
\begin{equation}
\nu'=\frac{\eta-\nu}{\eta-1}\,.
\end{equation}
Eq.~(\ref{eq:eigen-2}) can be transformed into the Schr\"odinger
equation \cite{Risken1989}
\begin{equation}
-\frac{1}{2}\frac{d^{2}}{dz^{2}}\psi_{\lambda}(z)+V(z)\psi_{\lambda}(z)=\lambda\psi_{\lambda}(z)\label{eq:schroed}
\end{equation}
with the potential
\begin{equation}
V(z)=\frac{1}{8z^{2}}\nu'(2+\nu')\,.
\end{equation}
Here $\psi_{\lambda}(z)=P_{\lambda}^{\prime}(z)/\sqrt{P_{0}^{\prime}(z)}$.
The condition of zero probability current at the reflective boundaries
$z=z_{\mathrm{min}}$ and $z=z_{\mathrm{max}}$ become 
\begin{equation}
\left.\left(\frac{d}{dz}+\frac{\nu'}{2}\frac{1}{z}\right)\psi_{\lambda}(z)\right|_{z=z_{\mathrm{min}},z_{\mathrm{max}}}=0\,.
\end{equation}
The solution of Eq.~(\ref{eq:schroed}) corresponding to the eigenvalue
$\lambda=0$ is
\begin{equation}
\psi_{0}(z)=\sqrt{\frac{\nu'-1}{z_{\mathrm{min}}^{1-\nu'}-z_{\mathrm{max}}^{1-\nu'}}}z^{-\frac{\nu'}{2}}\,.
\end{equation}
Eq.~(\ref{eq:schroed}) can be solved using standard finite-difference
or finite-element methods. Having the eigenfunction $\psi_{\lambda}(z)$
the first moment of the stochastic variable $x$ can be calculated
using the equation
\begin{equation}
X_{\lambda}=\int_{z_{\mathrm{min}}}^{z_{\mathrm{max}}}\psi_{0}(z)|\eta-1|^{\frac{1}{1-\eta}}z^{\frac{1}{1-\eta}}\psi_{\lambda}(z)\, dz\,.
\end{equation}

\begin{figure}
\includegraphics[width=0.6\textwidth]{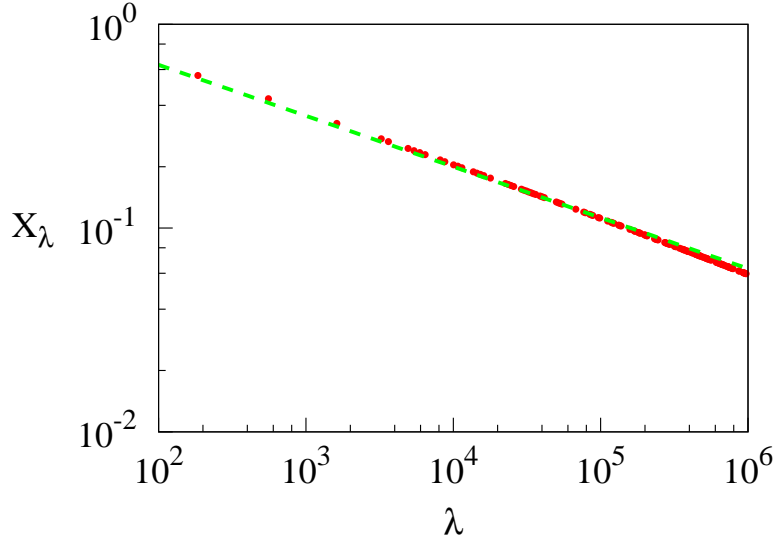}

\caption{(Color online) Dependence of numerically obtained first moments of
the variable $x$ on the eigenvalues $\lambda$ for the lowest eigenvalues
(red dots). Eigenvalues and eigenfunctions are obtained numerically
solving Eq.~(\ref{eq:schroed}). The dashed green line shows the
slope $\lambda^{-0.25}$, predicted by Eq.~(\ref{eq:x-lambda}).
The parameters used are $\eta=\frac{5}{2}$, $\nu=3$, $x_{\mathrm{min}}=1$
and $x_{\mathrm{max}}=1000$.}
\label{fig:eigen}

\end{figure}

Let us take the following values of the parameters in Eq.~(\ref{eq:FP-frac}):
$\eta=\frac{5}{2}$, $\nu=3$. The dependence of the numerically calculated
first moment $X_{\lambda}$ on the eigenvalue $\lambda$ for lowest
eigenvalues is shown in Fig.~\ref{fig:eigen}. We see a good agreement
with the analytical prediction (\ref{eq:x-lambda}) of power-law dependence
on $\lambda$. For larger eigenvalues $\lambda$ than those shown
in Fig.~\ref{fig:eigen} the power-law dependence does not hold and
$X_{\lambda}$ decrease faster.

\begin{figure}
\includegraphics[width=0.6\textwidth]{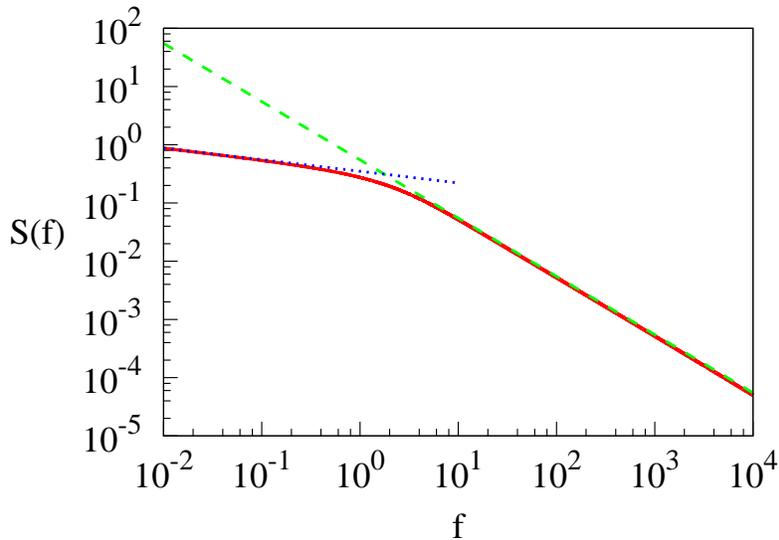}

\caption{(Color online) Power spectral density for the diffusion process defined
by Eq.~(\ref{eq:FP-frac}) with the parameter $\alpha=0.8$. The
solid red line shows the result of numerical calculation using Eq.~(\ref{eq:psd-gen}).
The dashed green line shows the slope $1/f$, whereas the dotted blue
line shows the slope $f^{-0.2}$. Other parameters are the same as
in Fig.~\ref{fig:eigen}.}
\label{fig:spectrum}

\end{figure}

The PSD calculated using Eq.~(\ref{eq:psd-gen}) is presented in
Fig.~\ref{fig:spectrum}. Eigenvalues $\lambda$ and the first moments
$X_{\lambda}$ shown in Fig.~(\ref{fig:eigen}) have been used. We
see a good agreement with the predicted power-law dependency of the
PSD on the frequency for frequencies $f>f_{\mathrm{min}}\approx1$.
The power-law exponent coincides with Eq.~(\ref{eq:beta}). For smaller
frequencies $f<1$ the PSD exhibits the power-law behavior (\ref{eq:psd-small})
with the exponent $1-\alpha$.

\section{Conclusions}

\label{sec:concl} In summary, we proposed Eq.~(\ref{eq:time-fract-FP})
describing the subdiffusion of particles in an inhomogeneous medium
that generalizes the previously obtained time-fractional Fokker-Planck
equation with the position-independent diffusion coefficient. Fokker-Planck
equation with the position-independent diffusion coefficient has been
used to model various phenomena such as ion channel gating \cite{Goychuk2006}
and the translocation dynamics of a polymer chain threaded through
a nanopore \cite{Dubbeldam2007}. Properties of such equations has
been studied extensively. In this paper we analyzed a more general
case when both drift and diffusion coefficients are position-dependent.
We hope that the present model can serve as a basis to study trapping
induced subdiffusion in complex inhomogeneous media.

We derived the analytical expression of power spectral density of
signals described by the one-dimensional time fractional Fokker\textendash{}Planck
equation in a more general case when diffusion coefficient depends
on the position. The general expression for the PSD (\ref{eq:psd-gen})
we applied to a particular case (\ref{eq:FP-frac}) when the drift
and diffusion coefficients have power-law dependence on the position.
The resulting PSD has a power-law form $S(f)\sim f^{-\beta}$ in a
wide range of frequencies, with the power-law exponent $\beta$ given
by Eq.~(\ref{eq:beta}). This approximate results is confirmed by
the numerical simulation (see Fig.~\ref{fig:spectrum}). Thus, according
to Eq.~(\ref{eq:beta}), time-fractional Fokker-Planck equation with
power-law coefficients yields the PSD with the power-law exponent
equal to or larger than 1 in a wide range of intermediate frequencies.
In contrast, the PSD for small frequencies has a power-law dependency
on the frequency in the form of $f^{-(1-\alpha)}$ even when the diffusion
coefficient depends on the position.

Since an analytical solution of time-fractional Fokker-Planck equation
can be obtained only in separate cases, there is a need of numerical
solution. For the numerical solution of the nonlinear equations, such
as those resulting in Eq.~(\ref{eq:FP-frac}), we propose to use
a variable time step that adapts to the coefficients in the equation.
Such a variable time step is equivalent to changing of the operational
time $\tau$ to the position-dependent operational time $\tau^{\prime}$.

\end{document}